\documentclass[twocolumn,showpacs,floatfix,preprintnumbers,amsmath,amssymb]{revtex4}

\usepackage{amsmath}
\usepackage{amssymb}
\usepackage{graphicx}
\usepackage{dcolumn}
\usepackage{bm}
\usepackage{psfrag}
\usepackage{rotating}
\usepackage{color}
\usepackage{wasysym}

\newcommand{\tn}[1]{\textnormal{#1}}

\begin{document}
\title{Collective motion due to individual escape and pursuit response}

\author{Pawel Romanczuk$^1$, Iain D. Couzin$^2$, Lutz Schimansky-Geier$^1$}
\email{romanczuk@physik.hu-berlin.de}
\affiliation{$^1$ 
Department of Physics, Humboldt University Berlin, Newtonstr. 15, 12489 Berlin, Germany,
$^2$ Department of Ecology and Evolutionary Biology, Princeton University, Princeton, New Jersey 08543}

\begin{abstract}
Recent studies suggest that non-cooperative behavior such as cannibalism may also be a driving mechanism of collective motion. Motivated by these novel results we introduce a simple model of Brownian particles interacting by pursuit and escape interactions. We show the onset of collective motion due to escape and pursuit response of individuals and demonstrate how experimentally accessible macroscopic observables depend strongly on the ratio of the escape and pursuit strength. We analyze the different impact of the escape and pursuit response on the motion statistics and determine the scaling of the migration speed with model parameters.
\end{abstract}
\pacs{Valid PACS appear here}

\maketitle
The emergence of collective motion of living organisms, such as exhibited by 
flocks of birds, bacterial colonies or insect swarms is an ubiquitous and 
fascinating self-organization phenomenon in nature, which still bears many 
open questions.      
A common explanation for the emergence of collective motion in a wide range of animals is that it serves as a protection mechanisms against predators.
Recent experimental results suggest a novel mechanism driven by cannibalism  
which may, surprisingly, facilitate collective motion in mass migrating 
insects  \cite{simpson_cannibal_2006,bazazi_collective_2008}. Insect swarms 
can extend over many kilometers and can have a devastating impact. Locusts, 
for example, can invade up to one fifth of the Earth's land surface and are 
estimated to affect the livelihood of one in ten people in the planet 
\cite{uvarov_grasshoppers_1977}.

The phenomenon of swarming in general has attracted scientists from a wide range 
of disciplines with different scopes and perspectives \cite{collection}. 
In recent years it also be became the focus of 
an increasing number of publications in the 
field of statistical physics, non-linear dynamics and pattern formation. 
These contributions enhanced significantly our understanding of collective 
motion in systems of self-propelled particles (SPP) by discovering universal 
scaling laws and phase-transition like behavior, and offered new stimuli to 
the theory of non-equilibrium systems 
\cite{vicsek_novel_1995,toner_flocks_1998,aditi_simha_hydrodynamic_2002,gregoire_onset_2004,erdmann_noise_2005,dorsogna_self-propelled_2006,chate_simple_2006,peruani_nonequilibrium_2006,aldana_phase_2007,romanczuk_beyond_2008}.
Recent examples of the ongoing research are a special journal issue dedicated exclusively to active motion and swarming \cite{erdmann_active_2008} or the work by Grossman \emph{et. al.} \cite{grossman_emergence_2008}, where the authors demonstrate the emergence of collective motion of SPP interacting via inelastic collisions.

In this letter we investigate a simple but generic model of individuals with escape and pursuit behavior 
which may be associated with cannibalism. We show the onset of collective motion   
in the absence of explicit velocity-alignment interaction, which to our knowledge is an integral ingredient of most related models. Our work demonstrates a strong dependence of macroscopic dynamics on the relative strength of individual escape and pursuit response. Directed 
translational motion in our model is a strictly collective (but not cooperative) behavior and may be therefore termed group-propulsion.
Our model offers a novel perspective on possible mechanisms of onset and persistence of collective motion and the resulting migration patterns in nature and represents an interesting example of pattern formation and phase-transitions in non-equilibrium systems.

We model an individual organism as an active Brownian particle in two dimensions ($d=2$) with an internal energy depot (see \cite{abp} for details). This additional degree of freedom describes the energy budget of individuals determined by their uptake of nutrients, internal dissipation to maintain body processes and conversion of energy into energy of motion.  It allows individuals in our model to increase their speed in reaction to external stimuli by conversion of internal energy into energy of motion. 
For simplicity we assume throughout this work that at all times there is a surplus of internal energy  which allows us to neglect the explicit treatment of the energy balance and focus on the spatial dynamics only. 

Each individual (particle) obeys the following Langevin dynamics:
\begin{equation}
\dot{\bf{r}}_i  =  {\bf v}_i,  \quad \dot{\bf{v}}_i  = -\gamma {\tn{v}}_i^{\alpha-1}{{\bf v}}_i+{\bf F}^s_{i}
	+\sqrt{2 D_v}\boldsymbol \xi_i \label{eq_dynamicv}
\end{equation}
The first term on the left hand side of the velocity equation (\ref{eq_dynamicv}) is a friction term with friction coefficient $\gamma$ and an arbitrary power dependence on velocity represented by $\alpha=1,2,3,\dots$. The response of individual $i$ to other individuals is described by an effective social force ${\bf F}^s_i$. The last term is a  non-correlated Gaussian random force with intensity $D_v$. A solitary individual (${\bf F}^s_i=0$) explores its environment by a continuous random walk, where the individual velocity statistics are determined by $\gamma$, $\alpha$ and $D_v$.     
The parameters are given in arbitrary time and space units $T$ and $X$: $[\gamma]=X^{1-\alpha}T^{\alpha-2}$, $[D_v]=X^2 T^{-3}$. 
	
The finite-size of individuals is taken into account by fully elastic hardcore collisions with a particle radius $R_{hc}$  (for details see \cite{brilliantov_kinetic_2004}).

Motivated by experimental observations \cite{bazazi_collective_2008}, we introduce the following response mechanisms:  
If approached from behind by another individual $j$ the focal individual $i$ increases its velocity away from it in order to prevent being attacked from behind. We refer to this behavior as \emph{escape} (e). 
If the focal individual "sees" another individual up-front moving away, it increases its velocity in the direction of the escaping individual. We refer to this behavior as \emph{pursuit} (p). 
No response in all other cases.
The response of an individual is determined the following decision algorithm: A) Is there another individual within my sensory range $l_s>R_{hc}$; B) If yes, is it in front or behind me, and C) does it come closer or does it move away.

Based on the above considerations we write ${\bf F}^s_i$ as a sum of an effective escape and an effective pursuit force:
${\bf F}^s_i={\bf f}^e_i+{\bf f}^p_i$
with
\begin{subequations}\label{eq_forces}
\begin{align}
{\bf f}^e_i  & = \frac{\chi_e}{N_e}  \sum_j \Delta{\bf v}_{ji}
\theta(l_s-r_{ji}) \theta(-{\bf v}_{i}{\bf \hat r}_{ji}) \theta(-{\bf v}_{ji}{\bf \hat r}_{ji}) \label{eq_f_e} \\
{\bf f}^p_i  & = \frac{\chi_p}{N_p} \sum_j  \Delta{\bf v}_{ji}
\theta(l_s-r_{ji}) \theta({\bf v}_{i}{\bf \hat r}_{ji}) \theta({\bf v}_{ji}{\bf \hat r}_{ji}) \label{eq_f_p}
\end{align}
\end{subequations}
where $\chi_{p,e}\geq0$ are the corresponding interaction strengths, $\Delta{\bf v}_{ji}=({\bf v}_{ji} {\bf \hat r}_{ji}){\bf \hat r}_{ji}$ is the relative velocity of particle $j$ with respect to particle $i$, with ${\bf v}_{ji}={\bf v}_j-{\bf v}_i$ and ${\bf \hat r}_{ji}=({\bf r}_{j}-{\bf r}_{i})/|{\bf r}_{j}-{\bf r}_{i}|$. The Heaviside functions $\theta$ reflect the conditions for the escape and the pursuit response. Both forces are normalized by the respective number of individuals which the $i$-th individual responses to: $N_e$ is the number of individuals which fulfill the escape response conditions and $N_p$ the corresponding number for the pursuit response. 

The symmetry of the introduced interaction is broken in several ways: 
the interaction acts only on one of the interacting particles (\emph{action}$\neq$\emph{reaction});
the interactions are direction selective - the particles distinguish between their front (${\bf v}_i \cdot {\bf r}_{ji}>0$) and their back (${\bf v}_i \cdot {\bf r}_{ji}<0$) and between approach (${\bf v}_{ji} \cdot {\bf r}_{ji}<0$) and escape (${\bf v}_{ji} \cdot {\bf r}_{ji}>0$);  
the strength of interaction to the front and back may be different ($\chi_{e}\neq \chi_{p}$).
The most important property of the interactions is their anti-dissipative nature with respect to  kinetic energy. Note that ${\bf F}^s_i$ leads only to acceleration of individuals and is analogous to the autocatalytic machanism proposed in Bazazi \emph{et al.} \cite{bazazi_collective_2008}.

Throughout this letter we will discuss our numerical results in terms of the rescaled density 
$\rho_s=N l_s^2/L^2$,
where $N$ is the total particle number, $l_s$ the interaction range and $L$ the size of the simulation domain. All simulation results were obtained with periodic boundary condition. We will restrict here to the case of moderate noise intensity $D_v<1$ and focus on the system behavior with changing density $\rho_s$, which may easily be controlled in experiments. Although the system shows interesting global behavior with increasing noise $D_v$, related to phase transitions known from systems of SPP \cite{vicsek_novel_1995,gregoire_onset_2004,aldana_phase_2007}, the corresponding analysis is beyond the scope of this letter and will be discussed in a forthcoming publication.  

\begin{figure}
\psfrag{r = 2.25}{$\rho_s =2.25$}
\psfrag{r = 1.25}{$\rho_s =1.25$}
\psfrag{r = 0.3}{$\rho_s =0.30$}
\psfrag{p}{p}
\psfrag{e}{e}
\psfrag{p+e}{p + e}
\includegraphics[width=0.8\linewidth]{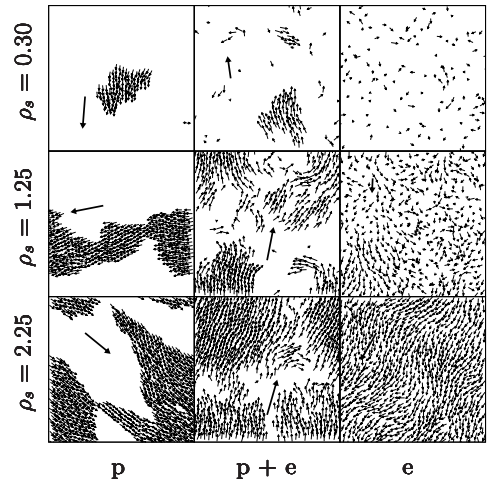}
\caption{Typical spatial configurations and particle velocities (small arrows) for pursuit only (p), pursuit+escape (p+e) and escape only (e) cases at different particle densities $\rho_s=0.30$, $1.25$, $2.25$. The direction of the large arrows indicates the mean migration direction and their length the migration speed $U$. For the escape only case at low densities the mean migration speed vanishes $U\approx0$. \label{fig_snapshot}}
\end{figure}

\begin{figure}
\includegraphics[width=\linewidth]{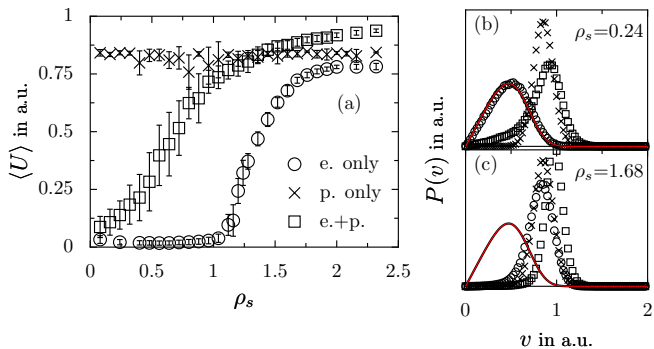}
\caption{(a) Mean global velocity $\langle U \rangle$ for escape-only ($\ocircle$) $\chi_e=10$, $\chi_p=0$, pursuit-only ($\times$) $\chi_e=0$, $\chi_p=10$ and symmetric escape+pursuit  ($\Square$) $\chi_e=\chi_p=10$ over particle density $\rho_s$ obtained from numerical simulations with periodic boundary conditions ($\gamma=1$, $D_v=0.05$, $\alpha=3$, $R_{hc}=1$, $l_{s}=4$). Only pure translational solutions were considered and the errorbars represent one std. deviation; Particle speed distribution $P(v)$ for the different interaction types in comparison with the analytical solution for non-interacting Brownian particles (solid line) at  $\rho_s=0.24$ (b) and $\rho_s=1.68$ (c). 
\label{fig_meanu}}
\end{figure}

Numerical simulation reveal that irrespective of the detailed model parameters, the pursuit and escape interactions lead to global collective motion at high particle densities $\rho_s$
and moderate noise intensities $D_v$ (Fig.~\ref{fig_snapshot}). At low  $\rho_s$ however we observe a very different behavior in dependence on the microscopic details of the model, where the velocity statistics and spatial migration patterns depend strongly on the relative strength of the escape and pursuit interaction $\chi_p$ and $\chi_{e}$. For  $\chi_{e}>0$ and $\chi_{p}\to 0$ with increasing $\rho_s$  a transition between a disordered state, with vanishing mean migration speed
$\langle  U \rangle=|\sum_i {\bf v}_i|/N=0$
and an ordered  state with $\langle U \rangle>0$ takes place. This resembles similar transitions reported for SPP with velocity alignment (Fig.~\ref{fig_snapshot},\ref{fig_meanu}) \cite{vicsek_novel_1995}. With increasing $\chi_{p}$ the transition shifts to lower $\rho_s$ until it vanishes. For $\chi_{p}>0$ and $\chi_{e}\to 0$ there is no dependence of $\langle U \rangle$ on $\rho_s$.
         
In order to understand the dynamics we investigate the influence of escape and pursuit  interactions independently, by analyzing the extreme cases: $\chi_{p}=0$, $\chi_{e}>0$ (only escape) and $\chi_{e}=0$, $\chi_{p}>0$ (only pursuit).

\begin{figure}
\begin{minipage}{0.28\linewidth}
\psfrag{(a)}{(a)}
\includegraphics[width=\linewidth]{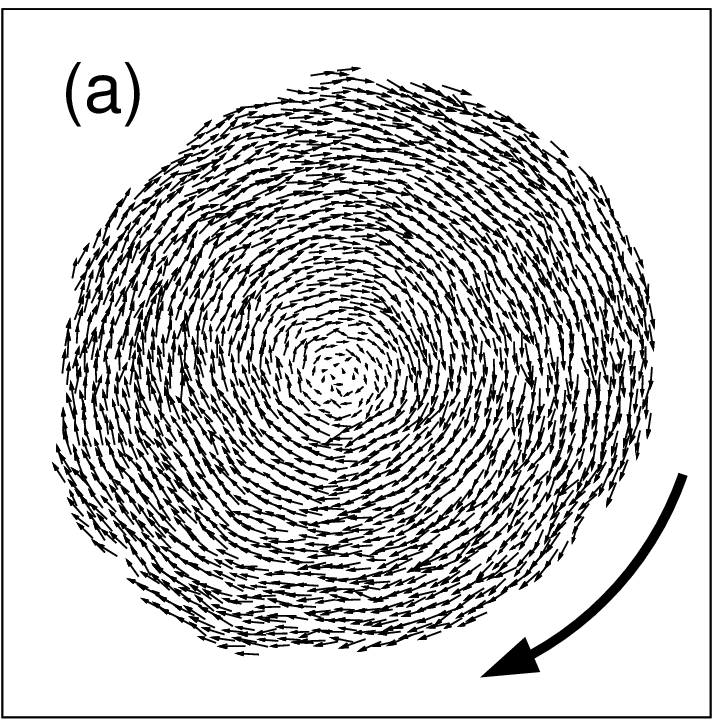}
\end{minipage}
\begin{minipage}{0.70\linewidth}
\includegraphics[width=\linewidth]{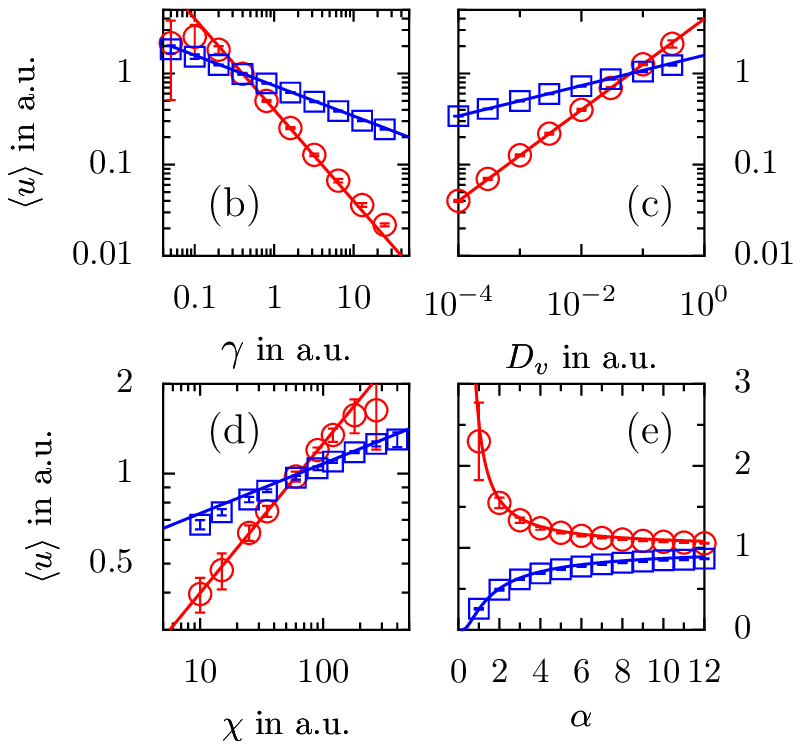}
\end{minipage}
\caption{(a) Snapshot of a single large vortex formed from random initial conditions for the pursuit only case. The arrow indicates the rotation direction; Comparison of numerically obtained average pair velocities $\langle u \rangle$ for $\alpha=1$ ($\Circle$) and $\alpha=3$ ($\Square)$ with the result of Eq.~\ref{eq_uspair} (solid lines): (b)  $\langle u \rangle$ over friction coefficient $\gamma$, (c) noise intensity $D_v$, (d) and interaction strength $\chi$. 
(e) $\langle u \rangle$ vs. friction function exponent $\alpha$. Here we distinguish two cases $A>1$ ($\Circle$) and $A<1$ ($\Square$), with $A=\sqrt{D_v\chi/\pi}/\gamma$.  \label{fig_cluster} }
\end{figure}

In the escape only case the particles try to keep their distance with respect to individuals approaching from behind. To the front only interactions via the hardcore collisions take place. At low $\rho_s$ after an escape response the probability of interaction within the characteristic time of velocity relaxation vanishes and the particles are able to reorient themselves (disordered state). As $\rho_s$ increases the frequency of escape interactions increases and the particles are able to correlate their velocities on several interaction length scales but subensembles may move in different directions. We observe a transition to the ordered state via an active fluid like state ($\rho_s\approx1.25$; Fig. \ref{fig_meanu}a). In the ordered state all particles are able to correlate their direction of motion. At all $\rho_s$ we obtain spatially homogeneous distribution of particles. The transition-like behavior is also reflected in the particle speed distribution $P(v)$. At low densities $P(v)$ corresponds to the analytical result obtained for non-interacting particles (Fig \ref{fig_meanu}b), whereas at high $\rho_s$ the maximum of the distribution shifts to higher speeds indicating a transition from pure random walk to directed translational motion (Fig \ref{fig_meanu}c).   

In the case of pursuit-only interaction the dynamics change dramatically. Already at low $\rho_s$ we observe a highly inhomogeneous state: initiated by formation of small compact particle clusters performing coherent translational motion. As there is no escape interaction the density of the clusters is only limited by the hardcore radius. At moderate noise intensities the clusters are highly stable and a process of cluster fusion can be observed where larger clusters absorb smaller clusters and solitary particles. The dominant stationary configuration with periodic boundary condition, and moderate noise, is a single large cluster performing translational motion (Fig.~\ref{fig_snapshot}). The migration speed $\langle U \rangle$ in Fig.~\ref{fig_meanu}a is given by the mean speed of a single cluster $\langle u \rangle=|\sum_{i\in \tn{cluster}} {\bf v}_i|/N_{\tn{cluster}} $, which for large clusters becomes independent of the cluster size and therefore independent of $\rho_s$. The same holds for $P(v)$ as shown in Fig. \ref{fig_meanu}b,c. An intriguing feature of pursuit-only is the possibility of the formation of large scale vortices out of random initial conditions due to collisions of clusters moving in opposite directions. After a nucleation a vortex may grow by absorbing smaller clusters leading to a single rotating structure (Fig.~\ref{fig_cluster}a) with life times exceeding $10^3$ time units. Preliminary results on vortex-stability indicate a monotonous increase of stability (i.e. life time) with size (not shown). The emergence of vortices in our model is in particur remarkable because so far they have only been reported for systems of SPP with confinement, or attracting potential, respectively \cite{czirok_formation_1996,dorsogna_self-propelled_2006, vollmer_vortex_2006}. Here the pursuit interaction acts in a sense as both: a propulsion mechanism and an asymmetric attraction.    

The analysis of the dynamics shows that both interactions --- escape and pursuit --- lead to collective motion but have an opposite impact on the density distribution. Whereas the escape interaction leads in general to a homogenization of density within the system, the pursuit interaction facilitates the formation of density inhomogeneities (clusters). This leads us to the insight that the actual escape+pursuit dynamics where $\chi_{p},\chi_{e}>0$ is a competition of the two opposite effects with respect on the impact on the particle density. The stability of moving clusters in this simple model is determined by the relative ratio of the interaction strengths. In general for the escape+pursuit case at low $\rho_s$ we observe fast formation of actively moving particle clusters with complex behavior: fusion and break up of clusters due to cluster collisions as well as spontaneous break up of clusters due to fluctuations. The weak dependence of the particle speed distribution $P(v)$ on $\rho_s$ combined with the clear deviation from the non-interacting case at low $\rho_s$ (Fig. \ref{fig_meanu}b,c) shows that the increase of $\langle U \rangle$ with $\rho_s$ for escape+pursuit originates in an alignment of individual cluster velocities.   

In order to determine the scaling of $\langle U \rangle$ with model parameters, 
we consider the smallest cluster which shows directed translational motion: a particle pair ($1,2$). We assume particle $1$ being in the front of particle $2$, $\chi_e=\chi_p=\chi$ and  $|{\bf r}_{12}|<l_s$  at all times. Through a transformation of Eq.~\ref{eq_dynamicv} into polar coordinates with ${\bf v}_{i}=(v_{i}\cos\varphi_{i},v_i\sin\varphi_i)$, where $\varphi_i$ is defined as the angle between ${\bf v}_i$ and $\bf \hat r_{12}$, it can be shown 
that for $-\pi/2<\varphi_i<\pi/2$ ($i=1,2$)   
the escape and pursuit interaction lead to an increase of either $v_1$ or $v_2$ in order to harmonize the speed of the slower particle with the faster one. The acceleration is counterbalanced by the frictional force and results in a non-vanishing translational velocity of the particle pair. In addition the interaction stabilizes the translational motion along ${\bf \hat r}_{12}$, i.e. $\langle \varphi_{i}\rangle \to 0$.
After the system relaxes to a stationary state (${\bf \hat r}_{12}$ varies slowly in time) we end up with effectively one-dimensional translational motion of the particle pair. 

The evolution of the mean speed of a particle pair in this one-dimensional situation $u_{1d}=(v_1+v_2)/2$ with $v_1\approx v_2$ can be approximated as
\begin{equation}
\frac{\Delta u_{1d}}{\Delta t} \approx -\gamma u_{1d}^\alpha + \frac{1}{2}\chi \langle |\delta v_i| \rangle_{1d} . \label{eq_utime}
\end{equation}
The second term on the right hand side of Eq.~\ref{eq_utime} accounts for the acceleration of the particle pair due to the escape+pursuit interaction with $\delta v_i=u-v_{i}$. The factor $1/2$ accounts for the fact that at a given time only one of the particles accelerates.
 
The deviations of individual particle speed from the mean speed result from the action of the non-correlated stochastic forces.  
We approximate the expectation value of the speed deviations $\langle |\delta v| \rangle_{1d}$, by considering the speed deviations as discrete increments taken from a Gaussian distribution with zero mean and variance $\sigma^2_{1d}=2 D_v \tau$ (Wiener process). Replacing  $\tau$ by the relaxation time of the interaction $\chi^{-1}$ yields: 
$\langle |\delta v| \rangle_{1d}=2\sqrt{D_v/\pi\chi}\label{eq_deltaw}$.
The stationary velocity of a particle pair can be calculated from (\ref{eq_utime}) to: 
\begin{equation}
 u^{s}_{1d} = \left(\frac{1}{\gamma}\sqrt{\frac{\chi D_v }{\pi}}\right)^{\frac{1}{\alpha}}. \label{eq_uspair}
\end{equation}
This result is in excellent agreement with numerical simulations of individual particle pairs (Fig.~\ref{fig_cluster}) for wide parameter ranges. The scaling in Eq. \ref{eq_uspair} agrees with the measurements of the average speed of large clusters at moderate $D_v$ and with results of a mean field approximation which will be discussed elsewhere.  

In summary, we have presented an individual based model for the kinematic description of large groups of individuals, where each individual responses to others in its local neighborhood by escape and pursuit behavior. The response itself is described by an effective social force acting on each individual and is motivated by recent experimental results on mass migrating insects.

We have shown the onset of collective motion due to the escape and pursuit interaction. The analysis of the model dynamics shows that the macroscopic behavior, which can be observed in experiments, such as migration speed vs. density or the spatial migration patterns depend strongly on the relative strength of the escape and the pursuit behavior of individuals. Furthermore we were able to obtain the right scaling of the migration speed with model parameters which is confirmed by numerical simulations.
 
On the one hand, recent experiments on marching insects suggest that escape dominates their marching behavior \cite{bazazi_collective_2008}. In this case our model predicts a phase transition like behavior of the  mean migration speed $\langle U \rangle$ in dependence on the density $\rho$, which is supported by previous results \cite{buhl_from_2006}. On the other hand, the coherent moving clusters and vortex structures observed in our model for the pursuit only case resemble observations of fish schools \cite{whitesides_self-assembly_2002, couzin_collective_2002} and suggest the relevance of our model to a wide range of swarming phenomena in nature.

This work was supported by the DFG through the collaborate research center Sfb555 ``Complex Nonlinear Processes". I.D. Couzin gratefully acknowledges support from a Searle Scholar Award (08-SPP-201) and a Darpa grant \#HR0011-05-1-0057 to Princeton University.

\end{document}